\documentstyle[aps]{revtex}

\begin{document}
\title{Finite Heat conduction in 2D Lattices}
\author{Lei Yang$^{1,2}$, Yang Kongqing$^2$}
\address{$^1${\it John v. Neumann Institute, Forschungszentrum Juelich, D-52425}\\
Juelich, Germany\\
$^2${\it Department of physics, Lanzhou University, Gansu 730000, China}}
\maketitle

\begin{abstract}
This paper gives a 2D hamonic lattices model with missing bond defects, when
the capacity ratio of defects is enough large, the temperature gradient can
be formed and the finite heat conduction is found in the model. The defects
in the 2D harmonic lattices impede the energy carriers free propagation, by
another words, the mean free paths of the energy carrier are relatively
short. The microscopic dynamics leads to the finite conduction in the model.

\end{abstract}

The study of heat conduction in models of insulating solids is a rather old
and debated problem, and the more general problem is one of understanding
the nonequilibrium energy current carrying state of a many body system. The
most of the work on heat conduction investigated the process of heat
transport in 1D lattices. The different models have been studied for
obtaining Fourier's law, several kinds of factors have been taken into
account in the models, such as the nonlinearity, on-site potentials, mass
disorder and {\it etc.} Then the typical 1D lattices Hamiltonian is

\begin{equation}
H=\sum_{i=1}^N\left[ \frac{p_i^2}{2m_i}+V\left( q_{i+1}-q_i\right) +U\left(
q_i\right) \right] ,
\end{equation}
where $m_i$ represents the mass of the $i\,$th particle, $V$ is the
potential energy of internal forces, and $U$ is an on-site potential. Based
on these studies, several sufficient or necessary conditions of the normal
thermal conductivity in a 1D lattices are suggested, such as
''nonintegrability is not sufficient to guarantee the normal thermal
conductivity in a 1D lattices'', ''in the Fourier law the phonon-lattice
interaction is the key factor in 1D on-site potential or mass disorder
lattice'', and recently Ref\cite{Momentum} proved rigorously that the
conductivity as given by the Green-Kubo formula always diverges in one
dimensional momentum conserving systems, Ref\cite{PeriodicPotential} and Ref%
\cite{Finite} give 1D models where momentum is conserved and yet the
conductivity is finite.

Several models have been studied on 2D lattices heat conduction, for
instant, in Ref\cite{Lorenctz 2d} a 2D Lorenctz gas, which describes a gas
of non-interacting point particles moving in a box, is presented, in Ref\cite
{Toda 2d} numerical simulations are performed for the 2D Toda-lattice. And
the divergence of the heat conductivity in the thermodynamic limit is
investigated in 2D lattices models of anharmonic solids with
nearest-neighbor interaction from single-well potentials by A.Lippi and
R.Livi\cite{FPU 2d}.

Since, investigating the property of thermal conductivity is in order to
understand that the macroscopic phenomena and their statistical properties
are in terms of deterministic microscopic dynamics. We can rough classify
the 1D lattices model in to two categories. The first category includes
homogeneous hamonic chains\cite{harm1d}, Toda lattices\cite{OnSite}(in the
models no temperature gradient can be formed), FPU lattices\cite{FPU 1d}
(the thermal conductivity $k\sim N^\alpha $ is divergent as one goes to the
thermodynamic limit N-%
\mbox{$>$}%
unlimit, and $\alpha $ is different in Toda lattices and FPU lattices) and
etc. The character of these models is that the freely propagating energy
carriers (particles, phonons or excitations) exists, then the finite heat
conduction do not exist in these models. The second category includes other
models, in which the free propagation energy carriers (particles, phonons or
excitations) can not be found, such as the ding-a-ling model\cite
{ding-a-ling} (where a set of particles harmonically anchored to an external
periodic lattices alternated with free particles), the ding-dong model\cite
{ding-dong} (a modification of the pervious system), the Frenkel-Kontorova
model\cite{OnSite} (where on-site potential is introduced), the 1D mass
disordered FPU lattices at low temperature\cite{FPUdisorder} and the models
in Ref\cite{PeriodicPotential}\cite{Finite}. Then the finite heat conduction
can be found in the models.

In solids, the local theory and the nonlocal theory\cite{nonlocal} of
thermal conductivity have been addressed for different systems. Undoubtedly,
the local theory has worked well in many applications, and nonlocal theory
is able to give explain some experiments\cite{nonlocal2} where the local
theory do not work. Which theory should be adopted in a special system, it
depends on the mean free path of the heat carriers, since energy is
transported by different kinds of heat carriers in the different systems. A
local theory is adequate when the mean free paths are relatively short, the
Fourier's thermal conductivity has always been described using a local
theory. When the mean free paths are relatively long, nonlocal heat
conductivity gives a better describes. For example, the sample size is
smaller than the mean free path, which corresponding to ballistic transport%
\cite{ballistic}, a nonlocal theory is adequate.

In this paper we discuss heat conduction in a 2D harmonic lattices. In Ref%
\cite{harm1d}, rigorous studies have shown that the thermal conductivity $%
\kappa $ diverges in 1D pure harmonic lattices. Afterwards, it was expected
that phonon waves should be damped by the scattering processes due to
impurity defects (disorder), thus the finite heat conduction should be
obtained. Unfortunately, it was found that isotopic disorder in a harmonic
chain yields a diverging conductivity\cite{harm1ddisorder}. These results
suggest that Fourier law can not be obtained with 1D harmonic chains. Then,
how are about 2D harmonic lattices? The present paper deals with a 2D
harmonic lattices with the missing bond defects, which is an other kinds
important defects and different form impurity defects(disorder), such as in
solid $H_2O$, there are the L defect (a missing hydrogen bond ), which play
also an important role in the ice properties related to transport and
relaxation\cite{ice}.

Consider a 2D lattices model as follow (Fig.1.): the 2D square lattices is
made of $N_x\times N_y$ particles, the equilibrium positions of the
particles labelled by the index $(i,j)$, every particle just interacts with
the nearest neighbor particle by the harmonic restore-force, and some
missing bond defects exist in the lattices, the vibration of every particle
restricts in one dimension. The Hamiltonian of the 2D lattices model with
defects is

\begin{equation}
H=\sum_{i=1,j=1}^{N_x,N_y}\frac{p_{ij}^2}{2m_{i,j}}+\sum_{i=1,j=1}^{N_x,N_y}%
\left[ \frac 12\alpha _{i,j}\left( q_{i,j+1}-q_{i,j}\right) ^2+\frac 12\beta
_{i,j}\left( q_{i+1,j}-q_{i,j}\right) ^2\right] .
\end{equation}
where $p_{i,j}$ is the momentum of the $(i,j)$ particle, $q_{i,j}$ is the
displacement from the equilibrium position. And $\alpha _{i,j}$, $\beta
_{i,j}$ are coefficients of the interactive force, here $\alpha _{i,j}$ and $%
\beta _{i,j}$ just take two kinds values $0$ or $1$, $\alpha _{i,j}=0$, $%
\beta _{i,j}=0$ means that a missing bond defect exists, here the capacity
ratio of defects is order to $\gamma $, $\gamma =0$ means that the lattices
is a pure hamonic lattices. Since the mass disorder do not be considered,
the dimensionless mass $m_{ij}$ is unity for the lattices. For simpler
expression, take $N_x=N_y$ and $N=N_xN_y$. The periodic boundary conditions
are assumed in the left and right boundary. The particles $(1,j)\ j=1,\ldots
,N_y$ in the top boundary of the lattices contact with the high temperature
heat bath $T_1$ and the particles $(N_x,j)\ j=1,\ldots ,N_y$ in the bottom
boundary contact with the low temperature heat bath $T_2$. And the
Nose-Hoover\cite{NoseHoover} heat bath act on the top line particles and the
bottom line particles, keeping them at temperature $T_1$ and $T_2$,
respectively. The equations of motion of these particles are determined by 
\begin{equation}
\begin{array}{l}
\stackrel{..}{q}_{1,j}=-\xi
_{1,j}p_{1,j}+q_{2,j}+q_{1,j+1}+q_{1,j-1}-4q_{1,j}, \\ 
\stackrel{.}{\xi }_{1,j}=\frac{p_{1,j}^2}{T_2}-1;\ \ \ \  \\ 
\stackrel{..}{q}_{N_x,j}=-\xi
_{N_x,j}p_{N_x,j}+q_{N_x-1,j}+q_{N_x,j+1}+q_{N_x,j-1}-4q_{N_x,j}, \\ 
\stackrel{.}{\xi }_{N_x,j}=\frac{p_{N_x,,j}^2}{T_1}-1,\ \ \ \ \ j=1,\ldots
,N_y,\ \ and\ j=0\ \Leftrightarrow \ j=N_y
\end{array}
\end{equation}
Two physical observables, the dynamical temperature and the heat flux of the
2D lattices model need to be defined. The definition of the local
temperature, which is same as the local temperature of 1D lattices system,
is the time average of the kinetic energy of the particle

\begin{equation}
T_{i,j}=\left\langle p_{ij}^2\right\rangle ,
\end{equation}
where $\left\langle \cdot \right\rangle $ denotes time average. The
components of the local heat flux vector $\overrightarrow{J}_{i,j}$ is

\begin{equation}
J_{i,j\rightarrow i,j+1}=\left\langle p_{i,j}f_{i,j+1}\right\rangle ;\
J_{i,j\rightarrow i+1,j}=\left\langle p_{i,j}f_{i+1,j}\right\rangle
\end{equation}
where $J_{i,j\rightarrow i,j+1}$ denoted to the flow of potential energy
form the particle $\left( i,j\right) $ to the particle $\left( i,j+1\right) $%
; $J_{i,j\rightarrow i,j+1}$ denoted to the flow of potential energy form
the particle $\left( i,j\right) $ to the particle $\left( i+1,j\right) $. It
is worth to define the total heat flux value of the lattices system,

\begin{equation}
N\left| {\it J}_{i,j}\right| =\sum_{i=1,j=1}^{N_x,N_y}\left| \overrightarrow{%
J}_{i,j}\right| =\sum_{i=1,j=1}^{N_x,N_y}\sqrt{\left( J_{i,j\rightarrow
i,j+1}\right) ^2+\left( J_{i,j\rightarrow i+1,j}\right) ^2}
\end{equation}
Then the time evolution of the displacement and the momentum of each
lattices point is calculated in system (1). All values and processes were
analyzed for time scales of $10^6-10^7$. The 2D lattices has been simulated
for the following capacity ratio of defects $\gamma =0.0$, $0.006$, $0.016$, 
$0.026$, $0.06$, $0.16$, $0.21$, $0.26$, $0.31$, $0.36$, $0.46$ and for the
following particle numbers $N=31^2$, $41^2$, $51^2$, $61^2$, $71^2$ and $%
111^2$. Additional numerical simulation shows the process of the wave
propagation in the lattices.

The numerical simulation of the 2D lattices has demonstrated that the
different value of $\gamma $ leads to very different heat conductivity. When 
$\gamma =0$, it means that the 2D model is pure harmonic lattices, and the
2D model provide similar results to 1D pure harmonic model, no temperature
gradient can be formed. When $\gamma $ is large enough, the temperature
gradient can be formed in the 2D model. In Fig.2, the temperature profiles
at $\gamma =0$ is plotted and the results is expected, the slice of the
figure at any fixed $j$ shows that the temperature profiles is as same as 1D
pure harmonic model. Increasing the capacity ratio of defects in the
lattices, the temperature gradient is formed. A typical example is shown in
Fig.3. Next figure (Fig.4) shows the distribution of the heat flux at same
parameters as Fig.3. It is obvious that the missing bond defects block the
heat flux's direct advance, and the heat flow is different at the every
lattices position.

Now the total particles number $N$ dependence of the total heat flux value $%
N\left| {\it J}_{i,j}\right| $ is shown in Fig.5. The total heat flux value
increase with particles number increase becomes slow, when increase the
capacity ratio of defects in the lattices. When $\gamma >0.26$, the total
heat flux value nearly is a constant with particles number increase. And we
calculate the larger lattices $N=111^2$, the numerical results suggest that
the finite heat conduction exists in the model.

Finally, the process of the wave propagation in the model is checked at $%
\gamma =0$ and $\gamma =0.26$. At time $t=0$, we give a excitations on the
boundary of the lattices(Fig.6), then the snapshot at some time $t_1$is
recorded. Fig.7 records the results of $\gamma =0$ (the pure harmonic
lattices), Fig.8 records the results of $\gamma =0.26$ (the finite
conduction lattices). Fig.7 shows that in the pure harmonic 2D lattices the
freely propagating energy carriers exists or the mean free paths are
relatively long. Fig.7 shows that in the harmonic 2D lattices with defects
the freely propagating energy carriers do not be found or the mean free
paths are relatively short. So, the harmonic 2D lattices with defects($%
\gamma =0.26$) exhibits the finite conduction..

Hence, the above results allow one to conclude: when $\gamma =0$, it means
that the 2D lattices is a pure harmonic lattices. The microscopic dynamics
shows that the freely propagating energy carriers exists or the mean free
paths are relatively long. The macroscopic phenomena is that the temperature
gradient can not be formed and the total heat flux value increase with
particles number increase. The 2D lattices lead to a infinite heat
conduction; when $\gamma >0.26$, it means that defects exist in the 2D
harmonic lattices. The microscopic dynamics shows that the freely
propagating energy carriers not be found or the mean free paths are
relatively short. The macroscopic phenomena is that the temperature gradient
can be formed and the total heat flux value do not change with particles
number increase. Then the 2D lattices lead to a finite heat conduction. In
summary, this paper give a hamonic model with missing bond defects, when the
capacity ratio of defects is enough large, the temperature gradient can be
formed and a finite heat conduction is found in this model.

\begin{center}
{\bf Figure Caption}
\end{center}

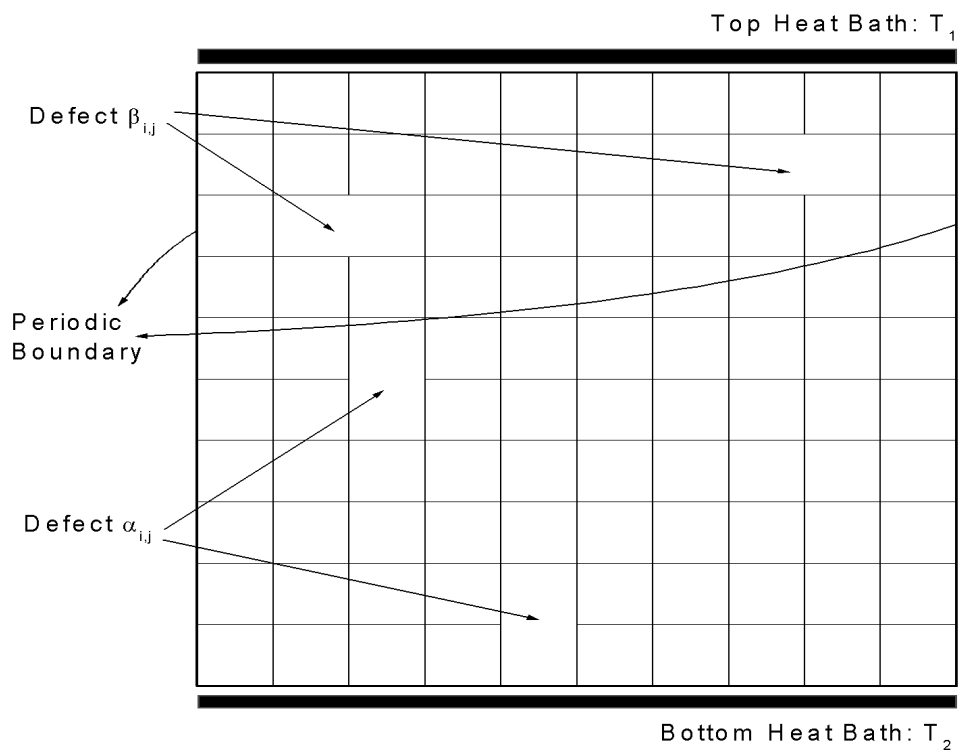
Fig.1. The 2D harmonic lattices model with missing bond defects is shown.

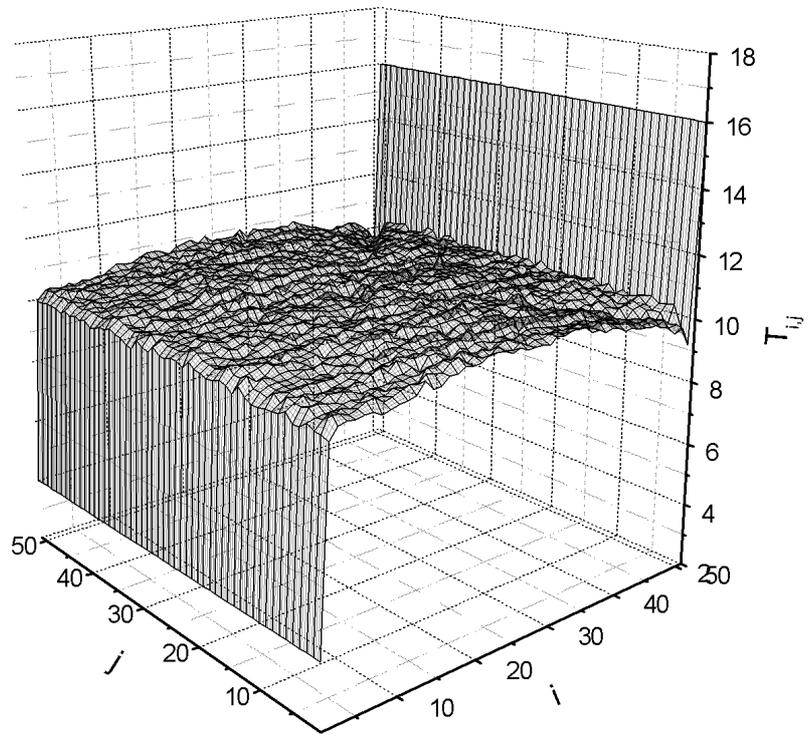
Fig.2. Temperature profile for the 2D pure hamonic model. The x axis and y
axis are the site index $(i,j)$ of the particle, the z axis is the dynamical
temperature $T_{i,j}$. The values of the parameters are as follow: $\gamma
=0.0$, the temperature of heat baths $T_1=16$, $T_2=4$ and system size $%
N=51^2$. The data are taken after the $10^6$ time interval.

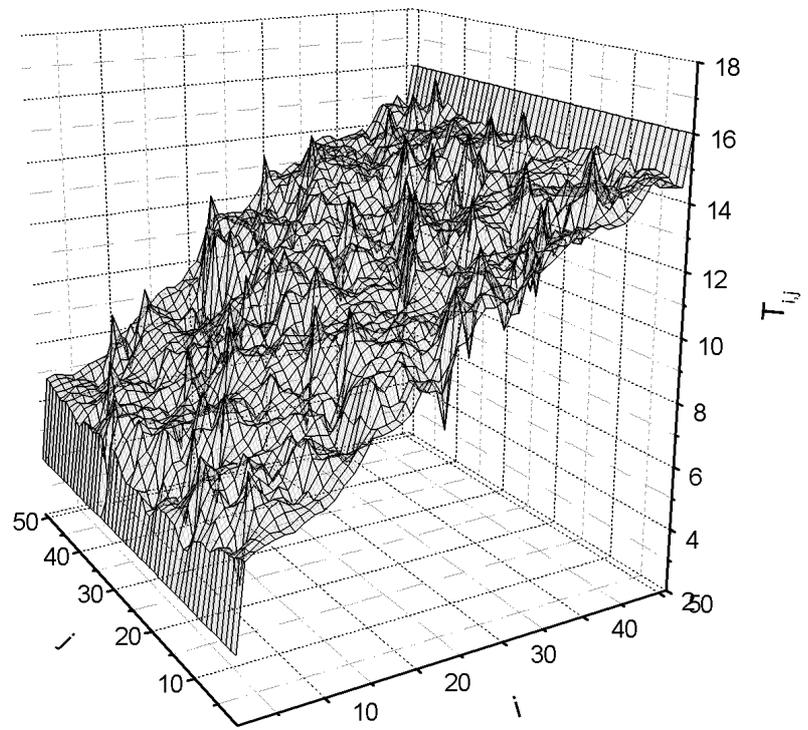
Fig.3. Temperature profile for the 2D hamonic lattices with missing bond
defects. The x axis and y axis are the site index $(i,j)$ of the particle,
the z axis is the dynamical temperature $T_{i,j}$. The values of the
parameters are as follow: $\gamma =0.06$, the temperature of heat baths $%
T_1=16$, $T_2=4$ and system size $N=51^2$. The data are taken after the $%
10^6 $ time interval.

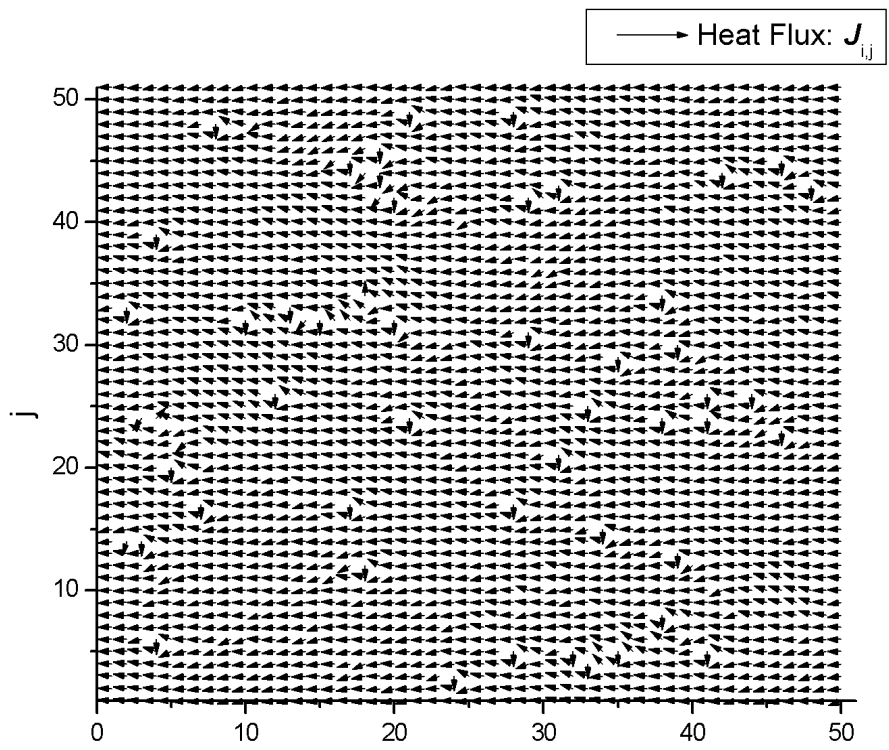
Fig.4. The stationary distribution of the heat flux vector $\overrightarrow{J%
}_{i,j}$ for the hamonic model with missing bond defects. The x axis and y
axis are the site index $(i,j)$ of the particle. The values of the
parameters are as follow: $\gamma =0.06$, the temperature of heat baths $%
T_1=16$, $T_2=4$ and system size $N=51^2$. The data are taken after the $%
10^6 $ time interval.

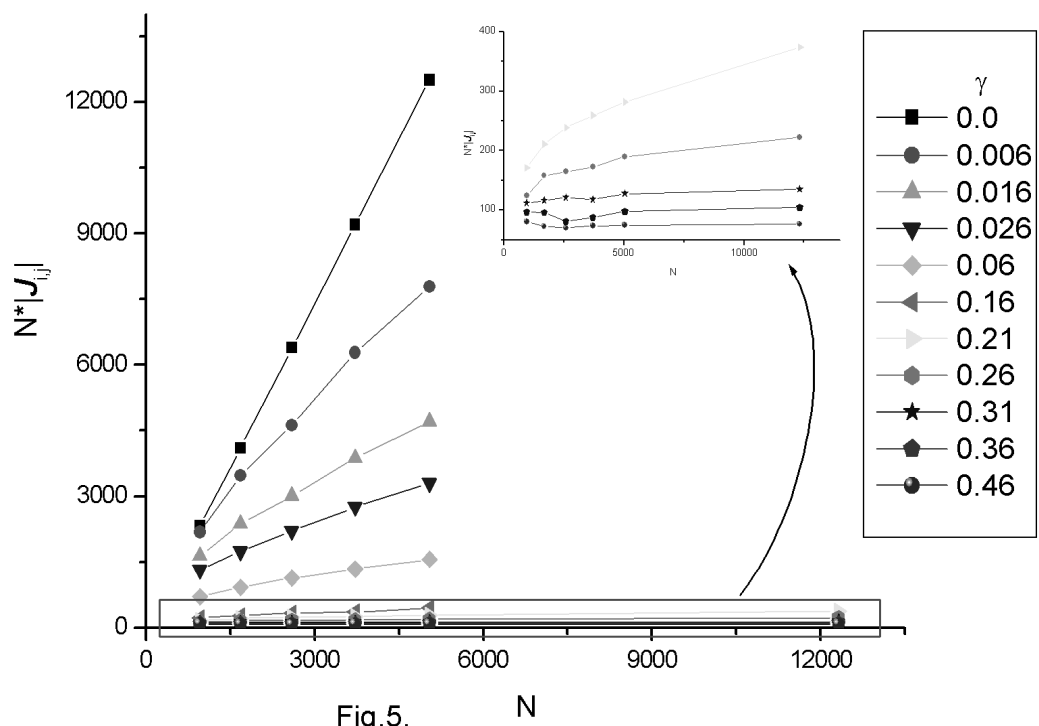
Fig.5. The total heat flux value vs the particles number at the same heat
baths temperature. The x axis is the particles number $N$, the y axis is
total heat flux value $N\left| {\it J}_{i,j}\right| $. The sub-figure
magnifies a part of the Fig.5 in the box of the red edge line, where $\gamma
=0.21$, $0.26$, $0.31$, $0.36$, $0.46$ and a additional particles number $%
N=111^2$.

Fig.6. Input pulse. The x axis and y axis are the site index $(i,j)$ of the
particle, the z axis is the energy $E$ of every particle. The system size is 
$N=111^2$. The figure plots the input pulse on the boundary.

Fig.7. Response to input pulse in the pure harmonic model. The x axis and y
axis are the site index $(i,j)$ of the particle, the z axis is the energy $E$
of every particle. The values of the parameters are as follow: $\gamma =0.0$%
, system size $N=111^2$. The figure shows the snapshot at $t=40$.

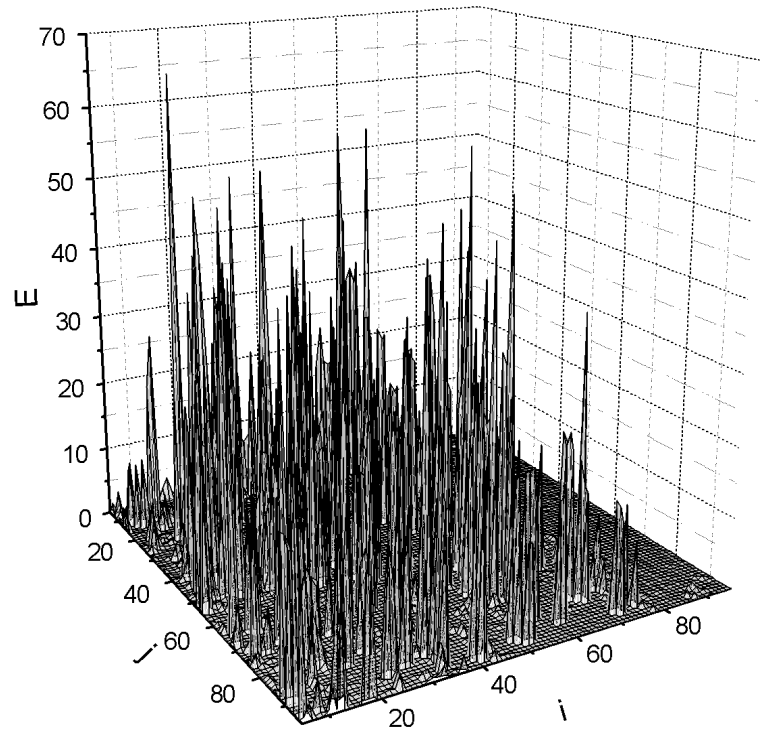
Fig.8. Response to input pulse in the harmonic model with defects. The x
axis and y axis are the site index $(i,j)$ of the particle, the z axis is
the energy $E$ of every particle. The values of the parameters are as
follow: $\gamma =0.26$, system size $N=111^2$. The figure shows the snapshot
at $t=40$.

\end{document}